\documentclass[letterpaper, 11pt]{article}
\pdfoutput=1
\usepackage[margin=1in]{geometry}
\usepackage{graphics}
\usepackage{array, multirow}
\usepackage{harvard}
\usepackage{pslatex}
\usepackage{paralist}
\usepackage{amsmath, amsthm, amssymb}
\usepackage{setspace}

\newtheorem{theorem}{Theorem}[section]

\newtheorem{proposition}[theorem]{Proposition}

\newtheorem{definition}[theorem]{Definition}

\begin{document}

\begin{titlepage}
\title{\sc A Dynamic Correlation Modelling Framework\\ with Consistent Stochastic Recovery}
\author{Yadong Li \thanks{
yadong.li@barcap.com. The views expressed in this paper are the author's own and may not necessarily reflect those of Barclays Capital.  The author is very grateful to Hongwei Cheng, Bjorn Flesaker, Sebastien Hitier, Ken Kyoung-Kuk Kim, Sam Morgan, Marco Naldi and Ariye Shater for many helpful comments and discussions. Part of this research is done in former Lehman Brothers.} \\
Barclays Capital \\
}
\date{First Version: February 26, 2009 \\ This Version: Apr 21, 2010}
\maketitle

\thispagestyle{empty}
\begin{abstract}
This paper describes a flexible and tractable bottom-up dynamic correlation modelling framework with a consistent stochastic recovery specification. The stochastic recovery specification only models the first two moments of the spot recovery rate as its higher moments have almost no contribution to the loss distribution and CDO tranche pricing. Observing that only the joint distribution of default indicators is needed to build the portfolio loss distribution, we propose a generic class of default indicator copulas to model CDO tranches, which can be easily calibrated to index tranche prices across multiple maturities.

This correlation modelling framework has the unique advantage that the joint distribution of default time and other dynamic properties of the model can be changed separately from the loss distribution and tranche prices. After calibrating the model to index tranche prices, existing top-down methods can be applied to the common factor process to construct very flexible systemic dynamics without changing the already calibrated tranche prices.  This modelling framework therefore combines the best features of the bottom-up and top-down models: it is fully consistent with all the single name market information and it admits very rich and flexible spread dynamics.

Numerical results from a non-parametric implementation of this modelling framework are also presented. The non-parametric implementation achieved fast and accurate calibration to the index tranches across multiple maturities even under extreme market conditions. A conditional Markov chain method is also proposed to construct the systemic dynamics, which supports an efficient lattice pricing method for dynamic spread instruments. We also showed how to price tranche options as an example of this fast lattice method.  

\end{abstract}

\end{titlepage}

\section{Introduction}
The base correlation model remains the most common method to price and risk manage synthetic CDOs \cite{bcexplained}. It is well known that the base correlation model is not arbitrage free, and it cannot produce a consistent joint default time distribution; therefore the base correlation model cannot be used to price and risk manage any default path-dependent or spread-dependent products. Not too long ago, the deterministic recovery assumption was the common practice within the base correlation framework.  However, in the recent market environments, models with the deterministic recovery often failed to calibrate to the index tranche market because it forces the senior most tranches to be risk free, leaving too much risk in the junior part of the capital structure.  \cite{rfl} first proposed the stochastic recovery for Gaussian Copula. More recently, a number of stochastic recovery specifications have been suggested for the base correlation framework, e.g. \cite{hitier} and \cite{krekel}. With these stochastic recovery specifications, the senior most tranches become risky, allowing the base correlation model to calibrate. However, most of the existing stochastic recovery specifications are not internally consistent, i.e., they can't be used to drive a Monte Carlo simulation and match the underlying CDS curves' expected recovery across time. The stochastic recovery specifications therefore introduced another source of inconsistency to the already inconsistent base correlation framework.

There have been a lot of efforts devoted to developing alternative models to the base correlation model in order to better price and risk manage the exotic correlation products whose payoff may depend on the default paths and tranche spreads.  One alternative modelling approach is to find a consistent static copula, which can produce the joint default time distribution in order to price default path-dependent instruments.  Random Factor Loading \cite{rfl} and the Implied Copula \cite{impliedcopula} \cite{skarke} are examples of the alternative static copulas.  Another alternative modelling approach is to develop dynamic correlation models, which can price the spread-dependent correlation instruments, e.g., tranche options.

There are two main categories of dynamic correlation models: the top-down approach and the bottom-up approach. The top-down approach directly models the dynamics of the portfolio loss distribution and ignores all the single name identities. The advantages of the top-down models include: 1) it is relatively easy to implement and calibrate and 2) it offers very rich spread dynamics.  The main disadvantages of the top-down models include: 1) it lacks the single name risk and sensitivity 2) it can't be used to price a bespoke CDO from the index tranches because the spread dispersion, which is a critical factor in CDO pricing, is not captured by the top-down models. \cite{shonbucher}, \cite{spa}, \cite{fwdloss}, \cite{kay} and \cite{bslp} are some representative examples of the top-down models. \cite{kay} and \cite{thin} also suggested the random thinning as a possible method to incorporate the single name information into the top-down approach.

The bottom-up approach, on the other hand, starts with the single name spread dynamics and a correlation structure; and then computes the portfolio and tranche spread dynamics as functions of the single name spread dynamics and the correlation structure. The advantage of the bottom-up approach is that it preserves the single name identities and the spread dispersion, and offers the single name sensitivity. A bottom-up model can produce the joint distribution of default times and spreads; therefore, it can cover a wider range of exotic correlation products than a top-down model. For example, any exotic contract whose payoff depends on the identity of an underlying issuer\footnote{For example, a vanilla bespoke CDO traded against a risky counterparty who does not post the full collateral. In this case, the identity of the counterparty is important.} cannot be easily handled with a top-down model.  However, a bottom-up model is much more difficult to implement and calibrate. Often, the model parameters that control the spread dynamics also affect the tranche prices; therefore the calibration to the index tranche prices can put severe restrictions on the resulting spread dynamics, making it difficult to produce the desired spread dynamics and fit quality to the index tranches simultaneously. Due to these difficulties, there is no known bottom-up model that can produce good index tranche calibration and flexible spread dynamics to the best knowledge of the author.  \cite{mortensen}, \cite{ml} and \cite{kogan} are some representative bottom-up dynamic correlation models. 

Under the current market conditions, the stochastic recovery is required for a bottom-up dynamic correlation model to achieve good calibration to the index tranche prices. Most of the existing stochastic recovery specifications cannot be directly used by a bottom-up dynamic correlation model because of their intrinsic inconsistencies. Defining a consistent and tractable stochastic recovery specification remains a challenge.

The paper is organized as follows: section \ref{sr} presents the consistent stochastic recovery specification; section \ref{frame} is the general framework of the dynamic correlation model; section \ref{imp} discusses the details of a non-parametric implementation of the general framework; section \ref{numres} shows the calibration and other numerical results of the non-parametric implementation; section \ref{dmc} proposes a conditional Markov chain extension and shows that the dynamic spread instruments can be priced efficiently using a lattice method.

\section{Consistent Stochastic Recovery \label{sr}}
This section first describes the generic properties of recovery rates; then proposes a tractable and consistent stochastic recovery specification. 

Define $\tau$ as the default time of an issuer, and ${\bf 1}_{\tau<t}$ as the indicator that the issuer defaults before time $t$.  The recovery rate $r(t_1, t_2)$ is a conditional random variable that represents the recovery rate conditioned on the issuer defaults between time $t_1$ and $t_2$, i.e.  $\tau \in (t_1, t_2)$.  $r(t,t)$ is used to denote the spot recovery rate when the issuer defaults exactly at time $t$, i.e., $\tau \in (t, t+dt)$.  

\begin{definition}
The following terms are defined for the recovery rate to simplify the exposition: 
\begin{compactenum}
\item spot mean: $\mu(t, t) = \mathbb{E}[r(t, t)]$
\item spot variance: $\sigma^2(t, t) = \textup{Var}[r(t, t)]$
\item term mean: $\mu(0, t) = \mathbb{E}[r(0, t)]$
\item term variance:  $\sigma^2(0, t) = \textup{Var}[r(0, t)]$
\end{compactenum}
\end{definition}

The term mean and variance of recovery rate are important for building the loss distribution at a given time horizon $t$ using the semi-analytical method \cite{semianalytical}. The spot mean and variance are useful inside a Monte Carlo simulation.  

\begin{proposition}
\label{recp}
The recovery rate has the following properties:
\begin{compactenum}
\item The recovery rate is positive and less than 1: $r(t_1, t_2) \in [0, 1]$, $\mu(t_1, t_2) \in [0, 1]$ 
\item The variance of the recovery rate is range bounded: $\sigma^2(t_1, t_2) \in [0, \mu(t_1, t_2)(1-\mu(t_1, t_2))]$ 
\end{compactenum}
\end{proposition}

The lower bound 0 of the recovery rate variance corresponds to the deterministic recovery. The upper bound of the recovery variance corresponds to a two point distribution with recovery rate values of $\{0, 1\}$ and probabilities of $\{1-\mu, \mu\}$, whose variance is the largest among all recovery distributions with mean $\mu$.

Consider two consecutive time periods of $(0, t_1)$ and $(t_1, t_2)$, the following equation holds because both sides are the recovery amount between time $(0, t_2)$:
\begin{equation}
\label{ind2p}
r(0, t_2) {\bf 1}_{\tau \in (0, t_2)}  =  r(0, t_1) {\bf 1}_{\tau \in (0, t_1)} + r(t_1, t_2) {\bf 1}_{\tau \in (t_1, t_2)} 
\end{equation}
Take the expectation on the previous equation:
\begin{equation}
\label{ind2pe}
p(t_2) \mu(0, t_2) = p(t_1) \mu(0, t_1) + (p(t_2) - p(t_1)) \mu(t_1, t_2) 
\end{equation}
where $p(t) = \mathbb{E}[{\bf 1}_{\tau < t}]$ is the default probability over time. Squaring both sides of 
(\ref{ind2p}), the cross term disappears because the two periods do not overlap, also note ${\bf 1}^2 = {\bf 1}$: 
\begin{equation*}
r^2(0, t_2) {\bf 1}_{\tau \in (0, t_2)}  =  r^2(0, t_1) {\bf 1}_{\tau \in (0, t_1)} + r^2(t_1, t_2) {\bf 1}_{\tau \in (t_1, t_2)} 
\end{equation*}
Then taking the expectation yields:
\begin{equation}
\label{ind2p2e}
p(t_2) \mathbb{E}[r^2(0, t_2)] = p(t_1) \mathbb{E}[r^2(0, t_1)] + (p(t_2) - p(t_1)) \mathbb{E}[r^2(t_1, t_2)] 
\end{equation}
Dividing the period between $(0, t)$ into infinitesimal time intervals, (\ref{ind2pe}) and (\ref{ind2p2e}) can be written in the following continuous form:

\begin{proposition}
\label{intr}
Suppose the default probability of the issuer $p(t) = \mathbb{E}[{\bf 1}_{\tau<t}]$ is continuous and differentiable with $t$.  The following relationship exists between the spot mean recovery $\mu(t, t)$ and the term mean recovery $\mu(0, t)$:
\begin{equation}
\label{int1}
\mu(0, t) = \frac{1}{p(t)}\int_0^t \mu(s, s)p'(s)ds = \frac{1}{p(t)}\int_0^{p(t)} \mu(p, p) dp
\end{equation}
It is always possible to write the $\mu(t,t)$ as $\mu(p,p)$ because the inverse 
function $t^{-1}(p)$ always exists since the $p(t)$ is monotonic and continuous. Similarly:
\begin{equation}
\label{int2}
\mu^2(0, t) + \sigma^2(0, t) = \frac{1}{p(t)}\int_0^t [\mu^2(s, s) + \sigma^2(s, s)] p'(s)ds 
= \frac{1}{p(t)}\int_0^{p(t)} [\mu^2(p, p) + \sigma^2(p, p)]  dp
\end{equation}
\end{proposition}

Note that the $\sigma^2(0, t)$ is not just an integration of the $\sigma^2(p,p)$, it also includes the contribution from changes in the $\mu(p,p)$.  An observation that immediately follows the Proposition \ref{intr} is that if the $\mu(p, p)$ and $\sigma^2(p, p)$ are chosen to be analytical functions of the default probability $p$, the $\mu(0, t)$ and $\sigma^2(0, t)$ can be computed just from the value of $p(t)$ at time $t$ using (\ref{int1}) and (\ref{int2}) regardless of the detailed shape of the $p(t)$ over time. This property is critical in developing the dynamic correlation modelling framework in the next section of this paper.

Considering a basket of $n$ credits indexed by the subscript $i = 1 ... n$, the notional amount of each credit is $w_i$. The portfolio loss at time $t$ is the sum of all the individual losses $L(t) = \sum_{i=1}^n w_i l_i$, where $l_i = {\bf 1}_{\tau_i<t} (1-r_i(0,t))$ is the loss for a unit notional amount of name $i$.  The mean and variance of the $l_i$ are easy to compute:
\begin{align*}
\mathbb{E}[l_i] &= p_i(t) (1-\mu_i(0,t)) \\
\textup{Var}[l_i] &= p_i(t) \sigma_i^2(0,t) + p_i(t) (1-p_i(t)) (1-\mu_i(0,t))^2 
\end{align*}
If the ${\bf 1}_{\tau_i<t}$ and $r_i(0,t)$ are independent between names, it is well known that the portfolio loss distribution at time $t$ can be approximated by a normal distribution according to the central limit theorem \cite{shelton}. The normal approximation to the loss distribution is fully characterized by the mean and variance of the portfolio loss $L(t)$, which can be computed as:
\begin{align}
\label{plmean}
\mathbb{E}[L(t)] &= \mathbb{E}[\sum_1^n w_i l_i] = \sum_1^n w_i \mathbb{E}[l_i] = \sum_1^n w_i p_i(t) [1-\mu_i(0,t)] \\
\label{plvar}
\textup{Var}[L(t)] &= \textup{Var}[\sum_1^n w_i l_i] = \sum_1^n w_i^2 \textup{Var}[l_i] = \sum_1^n w_i^2 p_i(t) [\sigma_i^2(0,t) + (1-p_i(t)) (1-\mu_i(0,t))^2] 
\end{align}
Therefore, the only recovery rate measures that are required to compute the loss distribution with the independent defaults and recovery rates are the $\mu_i(0, t)$ and $\sigma_i^2(0, t)$. The fine details of the recovery rate distribution other than the first two moments do not affect the portfolio loss distribution if $n$ is reasonably large so that the normal approximation is sufficiently accurate. The same argument can be made for any conditional independent correlation models, e.g., Gaussian Copula. 
\begin{proposition}
\label{recmv}
Given a conditional independent correlation model, the loss distribution at time $t$ is only sensitive to the first two moments of the term recovery distribution, i.e., $\mu_i(0, t), \sigma_i^2(0, t)$. The contribution of higher moments of the recovery rate is no more than the residual error of the normal approximation to the portfolio loss distribution.
\end{proposition}
The effects of the higher moments of the stochastic recovery distribution are quantified in section \ref{secmc} of this paper. Since the $\sigma_i^2(0,t)$ enters the variance of the portfolio loss in (\ref{plvar}), a stochastic recovery model has to specify both the mean and variance of the recovery rate in order to correctly reproduce the portfolio loss distributions over time.  Any stochastic recovery specification that does not capture the variance of recovery is inadequate by construction. Also, the stochastic recovery models that directly specify the term $\mu_i(0,t)$ and $\sigma_i^2(0,t)$, or the distribution for $r_i(0,t)$ are usually not consistent because the implied spot recovery $r_i(t,t)$ is not guaranteed to satisfy the constraints in the Proposition \ref{recp}. Most of the popular stochastic recovery specifications for the base correlation model, such as \cite{hitier} and \cite{krekel}, are not internally consistent for the reasons above.

In conclusion, a consistent and tractable stochastic recovery specification can be easily constructed by defining the analytical functions for the $\mu_i(p,p)$ and $\sigma_i^2(p,p)$. In a conditional independent model, the $\mu_i$ and $\sigma_i^2$ can be defined as functions of the conditional default probability. It is natural to choose the $\mu(p, p)$ to be a decreasing function, since it forces the recovery rates to be lower in the bad states of the economy when a lot of names default.  In a conditional independent model, the overall unconditional recovery rate is a weighted average of the conditional recovery rates over all possible states of the market factor.

Under this stochastic recovery specification, the expected recovery term structure is no longer constant.  The CDS curves are typically built with a constant recovery term structure, which is a convenient but arbitrary choice given that we don't observe the recovery term structure in the market. The recovery locks are only traded for distressed names at very short maturities, the bid/offer of single name recovery lock is often as large as 5-10\%. Therefore it is not a problem in practice to deviate from the constant expected recovery rate assumption as long as the single name default probabilities are bootstrapped accordingly so that the CDS contracts at all maturities are priced to the market.

Another advantage of this stochastic recovery specification is that it gives user some control of the recovery variance through the parameter $\sigma_i^2(p, p)$. The recovery variance is very important to the CDO tranche pricing and risk especially when a name is very close to default.

\section{Dynamic Correlation Modelling Framework\label{frame}}
\subsection{$JDDI(t)$ vs $JDDT$}
\cite{davidli} first introduced the default time copula to price multi-name credit derivatives. By definition, a default time copula and the marginal single name default time distributions fully determine the joint distribution of default times (abbreviated as $JDDT$). More formally, if $\tau_1, ... , \tau_N$ are the default times of the $N$ names in a credit portfolio, then $JDDT(t_1, ... , t_N) = \mathbb{P}\{\tau_1<t_1, ... , \tau_N<t_N\}$. Similarly, we can the joint distribution of default indicators for a given time horizon $t$ (abbreviated as $JDDI(t)$) as $JDDI(d_1, ... , d_N, t) = \mathbb{P}\{{\bf 1}_{\tau_1<t} = d_1, ... , {\bf 1}_{\tau_N<t} = d_N \}$, where the $d_i$ must be either 0 or 1. We use $\{JDDI(t)\}$ to denote a set of $JDDI(t)$ over a discrete sample of time $\{t\}$, which is usually the quarterly IMM dates for pricing synthetic CDOs. Both the $JDDT$ and $JDDI(t)$ are based on time 0 information $\mathcal{F}_0$ in the following discussion.

Since a default time $\tau$ describes the same event as a time series of default indicators that switches from 0 to 1 at $\tau$, the $JDDT$ can be viewed as the joint distribution of $N \times T$ default indicators where $N$ is the number of names in the portfolio and $T$ is the number of samples in time (which can be infinite for continuous time sampling). Given that the $JDDI(t)$ is the joint distribution of the $N$ default indicators at a given time $t$, therefore it is obvious that: 
\begin{proposition}
\label{marginal}
$\{JDDI(t)\}$ is the marginal distribution of the $JDDT$ at the given time grid $\{t\}$. Therefore, the $JDDT$ contains more information than $\{JDDI(t)\}$ and there can be infinitely many $JDDT$s having the same marginal distribution $\{JDDI(t)\}$. 
\end{proposition}

\begin{figure}
\caption{$JDDI(t)$ vs $JDDT$ \label{jddi}}
\center

\begin{minipage}{3.2in}
\center
\underline{Two $JDDT$s}
\vspace{.25cm}

\begin{tabular}{|cc|rr|}
\hline
{\bf $\tau_1$ Range} & {\bf $\tau_2$ Range} & {\bf $JDDT_1$} & {\bf $JDDT_2$} \\
\hline
     (2,$\infty$) &      (2,$\infty$) &       20\% &       20\% \\

     (2,$\infty$) &      (1,2) &        0\% &       10\% \\

     (1,2) &      (2,$\infty$) &        0\% &       10\% \\

     (1,2) &      (1,2) &       20\% &        0\% \\

     (2,$\infty$) &      (0,1) &       30\% &       20\% \\

     (1,2) &      (0,1) &        0\% &       10\% \\

     (0,1) &      (2,$\infty$) &       20\% &       10\% \\

     (0,1) &      (1,2) &        0\% &       10\% \\

     (0,1) &      (0,1) &       10\% &       10\% \\
\hline
\end{tabular}  
\end{minipage}
\begin{minipage}{3.2in}
\center
\underline{$\{JDDI(t)\}$}
\vspace{.25cm}

\begin{tabular}{|rr|rr|}
\hline
{\bf ${\bf 1}_{\tau_1<t}$} & {\bf ${\bf 1}_{\tau_2 < t}$} & {\bf $JDDI(t=1)$} & {\bf $JDDI(t=2)$} \\
\hline
         0 &          0 &       40\% &       20\% \\

         0 &          1 &       30\% &       30\% \\

         1 &          0 &       20\% &       20\% \\

         1 &          1 &       10\% &       30\% \\
\hline
\end{tabular}  
\end{minipage}

\end{figure}

To illustrate the relationship between the $\{JDDI(t)\}$ and the $JDDT$, Figure \ref{jddi} (left) showed two $JDDT$s for a portfolio with two names over two time periods. The two $JDDT$s have the identical marginal distribution $\{JDDI(t)\}$, which is shown on the right. Figure \ref{jddi} clearly shown that the $JDDT$ contains more information than the $\{JDDI(t)\}$. For example, if we consider an instrument that pays \$1 only if both name default within the time period (1, 2), its price can be uniquely determined by either of the $JDDT$s, but not by the $\{JDDI(t)\}$. It is also interesting to note that the two $JDDT$s in Figure \ref{jddi} produce different prices for this instrument even though their marginal distributions $\{JDDI(t)\}$ are identical.

Since the pioneering work of \cite{davidli}, it becomes a very common practice to calibrate a default time copula to index tranches, then use the calibrated default time copula to price other vanilla and exotic instruments. This common practice is seriously flawed because of the following key observation: 
\begin{proposition} 
\label{vanilla}
All the vanilla credit derivatives, such as CDS, CDO, NTD basket or CDO$^n$, can be priced from their expected survival (or loss) curves over time. Therefore their prices are fully determined by the $\{JDDI(t)\}$ and the stochastic recovery specification. 
\end{proposition}
Proposition \ref{vanilla} implies that the index tranche prices contain no information beyond $\{JDDI(t)\}$; therefore the $JDDT$ from a default time copula calibrated only to index tranche prices is completely arbitrary and it does not reflect any market information. It is dangerous to use such a default time copula to price exotic instruments whose value is sensitive to the $JDDT$. A more sensible modelling approach is to calibrate both the $\{JDDI(t)\}$ and $JDDT$ to relevant market observables: the $\{JDDI(t)\}$ of the model should be calibrated to the index tranche prices; while the $JDDT$ should be calibrated to exotic instruments whose value depends on the $JDDT$, such as forward starting tranches or tranche options. Ideally, the calibration of $\{JDDI(t)\}$ and $JDDT$ should not depend on each other so that their calibration can be carried out independently. In this section, we present a modelling framework with these properties.

\subsection{Default Indicator Copulas\label{condfun}}
The most fundamental building block of our dynamic correlation modelling framework is the copula functions on default indicators, which specifies the $\{JDDI(t)\}$ rather than the $JDDT$. Since a $JDDI(t)$ is specific to a given time horizon, default indicator copula functions have to be defined for every discrete sample of time in $\{t\}$ as oppose to the case in default time copula where a single copula function governs the dependencies across all time horizon. The $\{JDDI(t)\}$ of a credit portfolio has to satisfy the following constraint since the default event is irreversible: 
\begin{proposition}
\label{constraint}
In a credit portfolio, the probability of any subset of names being in the default state together has to monotonically increase over time. 
\end{proposition}
The following three conditions define a consistent set of default indicator copulas over time whose $\{JDDI(t)\}$ satisfy the constraint in 
Proposition \ref{constraint} by construction: 
\begin{definition}
\label{copi}
A set of default indicator copula functions over time can be defined by:
\begin{compactenum}
\item An increasing stochastic process $X_t$ that represents the common factor. The distribution function of the $X_t$ is denoted as $F(x,t) = \mathbb{P}\{X_t < x\}$, which is also referred to as the marginal distribution of $X_t$. $f(x, t)$ denotes the distribution density function of $X_t$: $f(x, t) = \frac{\partial F(x, t)}{\partial x}$. An increasing $X_t$ implies that: 
\begin{equation}
\label{phiz}
\frac{\partial F(x,t)}{\partial t} \le 0 \\
\end{equation}
\item A conditional default probability function $p_i(x, t) = \mathbb{E}[{\bf 1}_{\tau_i<t}|X_t=x]$ that satisfies the following constraints:
\begin{align}
\label{p01}
p_i(x,t) &\in [0, 1] \\
\label{pcon}
p_i(t) &= \mathbb{E}[p_i(x,t)] = \int p_i(x, t) f(x, t) dx \\
\label{pzz}
\frac{\partial p_i(x,t)}{\partial x} &\ge 0 \\
\label{pzt}
\frac{\partial p_i(x,t)}{\partial t} &\ge 0 
\end{align}
The $p_i(t)$ is the unconditional default probability for the i-th name, which is extracted from the underlying CDS curves. The $p_i(x,t)$ function needs to have some name specific parameters so that it can be calibrated to $p_i(t)$ according to (\ref{pcon}). Whenever $F(x,t)$ changes, the $p_i(x,t)$ has to be re-calibrated to $p_i(t)$ to ensure consistency with the underlying single CDS. (\ref{pzz}) and (\ref{pzt}) ensure that the conditional default probability $p_i(x,t)$ are increasing for any possible path of $X_t$ given that the $X_t$ is increasing.  
\item Default indicators ${\bf 1}_{\tau_i<t}$ are independent conditioned on $X_t = x$. 
\end{compactenum}
\end{definition}

In practice, any $p_i(x, t)$ function that satisfies the constraints (\ref{p01}) to (\ref{pzt}) can be used to construct the copula functions of default indicators. For example, a Gaussian default indicator copula can be constructed from Definition \ref{copi} by choosing:
\begin{align}
\label{gcjddi}
F(x, t) &= \Phi(x) \notag \\
p_i(x, t) &= \Phi(\frac{\Phi^{-1}(p_i(t)) - \sqrt{\rho} x}{\sqrt{1-\rho}})
\end{align}
where $\rho \in [0, 1)$ is the correlation and $\Phi(x)$ is the cumulative normal distribution function. In the Gaussian default indicator copula, the common factor $X_t$ is a constant process whose value is determined immediately after $t=0$, therefore the (\ref{phiz}), (\ref{pzz}) and (\ref{pzt}) are trivially satisfied. Even though Gaussian Copula was introduced by \cite{davidli} as a default time copula, we actually only need the Gaussian default indicator copula as in (\ref{gcjddi}) to price CDO tranches. The classic Gaussian copula lacks the degree of freedom to calibrate to index tranche prices, a more flexible specification of default indicator copulas is given in section \ref{imp}. 

Following \cite{semianalytical}, a CDO tranches can be priced semi-analytically under Definition \ref{copi} because of the conditional independence of the default indicators. We rewrite the mean and variance of portfolio loss conditioned on $X_t = x$ from \eqref{plmean} and \eqref{plvar} as:
\begin{align*}
\mu_L(x) &= \sum_1^n w_i p_i(x, t) [1-\mu_i(0, p_i(x, t))] \\
\sigma^2_L(x) &= \sum_1^n w_i^2 p_i(t) [\sigma_i^2(0,p_i(x, t)) + (1-p_i(x, t)) (1-\mu_i(0,p_i(x,t)))^2]
\end{align*}
Then the conditional ETL for a 0 to $K$ base tranche can be computed from the normal approximation:
\begin{equation}
\label{condetl}
\mathbb{E}[\min(L(t), K) | x] = K + (\mu_L(x) - K) \Phi(\frac{K-\mu_L(x)}{\sigma_L(x)}) 
- \sigma_L(x) \phi(\frac{K-\mu_L(x)}{\sigma_L(x)}) 
\end{equation}
where $\phi$ is the normal distribution density function. The conditional ETL can then be integrated over the $f(x, t)$ to obtain the unconditional ETL:
\begin{equation}
\label{etl}
\mathbb{E}[\min(L(t), K)] = \int \mathbb{E}[\min(L(t), K) | x] f(x, t) dx \\
\end{equation}
The ETL of a regular tranche with non-zero attachment is just the difference between the ETLs of two base tranches. To price super senior tranches, we also need the expected tranche amortization (ETA) due to default recovery, which can be computed in a similar manner as the ETL. A CDO tranche can be priced as a regular CDS once the ETL and ETA curves are known.  The semi-analytical pricing method is extremely fast with the normal approximation, and we will show later that it is accurate enough in practice.

\subsection{Model Dynamics and Time Locality}

Under Definition \ref{copi}, conditioned on a full path of $X_t$ sampled at time grid $\{t\}$, the default indicators at each time in $\{t\}$ are independent, which is equivalent to the independence of default times sampled at the same time grid. Therefore:
\begin{proposition}
\label{indt}
If the default time $\tau_i$ is discretely sampled at the same time grid $\{t\}$ as the common factor $X_t$, the $\tau_i$ are independent conditioned on the full path of $X_t$.
\end{proposition}

Definition \ref{copi} not only specified the default indicator copulas and the $\{JDDI(t)\}$, it also determines the $JDDT$ and the full systemic spread dynamics if more information on the dynamics of the common factor process is known:
\begin{proposition}
In Definition \ref{copi}, the dynamics of $X_t$ determines the systemic dynamics in the following way:
\begin{compactenum}
\label{info}
\item At any time $t$, the marginal distributions $F(x, t)$ determines the $JDDI(t)$. 
\item The Markov chain of $X_t$, ie: $\mathbb{P}[X_{t} | X_{s}]$ for all $\{t, s; t>s\}$, determines the $JDDT$.
\item The full dynamics of $X_t$ determines the joint distribution of default time and the systemic factors ($JDDTSF$)
\end{compactenum}
\end{proposition}
The first property in Proposition \ref{info} is due to the conditional independence of the default indicators, which ensures that $JDDI(t)$ is unique conditioned on a given value of $X_t$; therefore the distribution $F(x, t)$ fully determines the unconditional $JDDI(t)$. Similarly, the conditional independence of default time from Proposition \ref{indt} ensures that the $JDDT$ is unique conditioned on a given path of $X_t$, therefore the unconditional $JDDT$ is fully determined by a Markov chain on $X_t$ which specifies the distribution of all the possible paths of $X_t$. 

Proposition \ref{info} implies that each Markov chain on $X_t$ uniquely defines a default time copula. However, the Markov chain does not fully specify the dynamics of $X_t$ because the $X_t$ may further depends on other state variables, and there can be infinitely many different $X_t$ dynamics having an identical Markov chain. Therefore, the $JDDT$ can be viewed as the marginal distribution of the even broader $JDDTSF$, which is the joint distribution of default times, $X_t$ and other systemic factors. The full specification of $X_t$ dynamics has to uniquely determine the $JDDTSF$ because the $X_t$ is the only source of systemic randomness in Definition \ref{copi}.  

A key benefit that directly follows the first property in Proposition \ref{info} is the ``time locality'': if we change $F(x,t)$ at a given $t$, it only affects the $JDDI(t)$ for that time and it won't affect the $JDDI(t)$ at any other time. Furthermore, the $\mu_i(0, p_i(x,t))$ and $\sigma_i^2(0, p_i(x,t))$ of recovery rates in (\ref{int1}), (\ref{int2}) only depend on the conditional default probability $p_i(x,t)$, therefore: 
\begin{proposition}
The loss distribution from Definition \ref{copi} and the stochastic recovery specifications in (\ref{int1}), (\ref{int2}) also has the ``time locality'', i.e.: the loss distribution at a given time horizon $t$ in this modelling framework is fully determined by the marginal distribution $F(x, t)$. 
\end{proposition}
The ``time locality'' is a very important property that greatly simplifies the pricing and calibration across multiple maturities. With ``time locality'', the CDO tranches can be priced from the marginal distributions of $F(x, t)$ via \eqref{condetl}, \eqref{etl} without knowing the joint distribution (aka, the Markov chain) of $X_t$ across multiple time horizons; and the calibration to different maturity can be carried out almost independently by finding the appropriate marginal distribution $F(x,t)$\footnote{The $p_i(x,t)$ has to be re-calibrated using (\ref{pcon}) when changing the $F(x,t)$ in order to maintain the consistency with the single name default probability $p_i(t)$}. The only constraints on the calibration across maturities are the monotonicity constraint from (\ref{pzz}), (\ref{pzt}) and (\ref{phiz}), which are technical in nature and normally do not pose any serious limitations.  In contrast, the pricing and calibration across multiple maturities can be quite challenging in existing bottom-up models without the ``time locality'' property. For example, in the ``chaining'' model suggested by \cite{sidchain}, the tranche pricing requires multi-dimensional integration over the joint distribution of $X_t$ across multiple time horizons, which quickly become intractable numerically when the number of maturities increases.

Pricing and calibrating across multiple maturities consistently has been one of the most difficult modelling problem in synthetic CDOs.  We addressed this difficult problem effectively by constructing the modelling framework with the ``time locality'' property. The ``time locality'' property is a key consideration in the specification of the default indicator copulas in Definition \ref{copi}, as well as the stochastic recovery in (\ref{int1}) (\ref{int2}). The proposed modelling framework is a one-factor model with conditional independence as the $X_t$ is the only systemic factor. Even though this model does not explicitly capture the contagion effect, it an produce strong default clustering via large jumps in $X_t$. 

\subsection{Progressive Calibration}
Proposition \ref{info} directly connects the dynamics of $X_t$ to the systemic dynamics of the model. It allows us to calibrate the model progressively by choosing the appropriate properties of the $X_t$ process. For example, we can change the $JDDT$ without affecting the $\{JDDI(t)\}$ and the loss distribution by building different Markov chains on the same marginal distribution of $F(x, t)$; we can also change the $JDDTSF$ without changing the $JDDT$ by choosing different dynamics of $X_t$ while preserving the Markov chain of $X_t$. 

A very rich set of research has been published on building top-down models on the portfolio loss process. A very convenient observation is that if we add the additional constraint that the $X_t$ is positive, then the $X_t$ has the exactly the same properties as the portfolio loss process, i.e., they are both positive and increasing. Therefore, existing top-down methods that were intended for the portfolio loss process can be easily applied to construct the Markov chain or the full dynamics process of $X_t$. This hybrid approach combines the best features of the top-down and bottom-up models as it preserves all the single name information, and it admits very rich and flexible systemic spread dynamics (i.e.: the $JDDT$ and $JDDTSF$).

The copula functions in the Definition \ref{copi} also admit idiosyncratic spread dynamics. By ``idiosyncratic'', we mean factors that are only specific to a given issuer, which are independent from systemic factors as well as idiosyncratic factors of other issuers. The $p_i(X_0,t)$ term structure defines the default probability from the idiosyncratic spread dynamics because there is no contribution from systemic factors if $X_t$ remains constant at its initial value $X_0$.  The idiosyncratic spread dynamics could affect the pricing of certain exotic instruments, e.g., junior tranche options.

\begin{figure}
\caption{Progressive Calibration of the Model \label{calprog}}
\small
\center
\begin{tabular}{|c|m{2.2cm}|m{3cm}|m{3.5cm}|m{4cm}|}
\hline
{\bf Steps} & {\bf Model Info.} & {\bf Model Parameters} & {\bf Market Input}  & {\bf Products Covered} \\
\hline
         1 & $\{JDDI(t)\}$ & $F(x,t)$ & Single name CDS and Index Tranches, Very Liquid & Bespoke CDOs, NTD Basket, long/short CDO and CDO$^n$\\
\hline
         2 & $JDDT$ & Markov chain on $X_t$ & Some market observables on default path dependent instruments. Illiquid & All default path-dependent instruments, such as waterfall synthetics, forward starting or step-up tranches, loss triggered LSS \\
\hline
         3 & $JDDTSF$ & Full dynamics of $X_t$ & Very few market observables on tranche options, almost no liquidity & Products that depend on systemic spread dynamics: such as senior tranche options, spread triggered LSS etc \\ \hline
         4 & $JDDTSF$ + idiosyncratic spread dynamics & Full dynamics in $X_t$ + idiosyncratic dynamics compatible with $p_i(x, t)$ & Some market observables on single name swaption, some liquidity & Products that depend on both systemic and idiosyncratic spread dynamics, such as junior tranche options, etc. \\
\hline
\end{tabular}  
\end{figure}

Figure \ref{calprog} is a summary of the progressive calibration procedure made possible by Proposition \ref{info}.  In the progressive calibration procedure, each step only specify the necessary properties of the $X_t$ process to accommodate the corresponding market information. The earlier steps do not limit the generality of the later steps; and the later steps always preserve all the model parameters and properties from the earlier steps. The progressive calibration procedure in Figure \ref{calprog} is very attractive in practice because it allows instruments to be priced from the most liquid and reliable market information. For example, if the model is calibrated to the step 2 and is used to risk manage a book containing bespoke CDOs and loss triggered LSS, then we are certain that the bespoke CDO prices are fully determined by the liquid index tranches and underlying CDS curves; and they are not affected by the views or observations on the forward losses which may be used to calibrate the step 2.  Suppose there is new market information on the forward losses, then only step 2 of the model calibration needs to be updated, which only affects the pricing of the loss triggered LSS.

\section{A Non-parametric Implementation\label{imp}}
Following the general principles of the modelling framework presented in previous sections, we discuss the details of a non-parametric implementation of the model, and show how exotic instruments can be priced. The marginal distributions of $X_t$ in this non-parametric implementation are sampled discretely by a fixed grid in $\{x_j\}$. 

\subsection{Choosing the $p_i(x, t)$}
This section describes the details of a non-parametric implementation of this modelling framework, where the $p_i(x,t)$ function in the Definition \ref{copi} is chosen to follow that of \cite{ml}:
\begin{equation}
\label{pxt}
p_i(x,t) = 1 - c_i(t) e^{-\beta_i(t) x}
\end{equation}
The $\beta_i(t) \ge 0$ is a loading factor on the systemic process. For simplicity, $\beta_i(t)$ is chosen so that the systemic process contributes a constant fraction to the cumulative hazard:
\begin{equation}
\label{hzdsplit}
\log(\mathbb{E}[e^{-\beta_i(t) x}]) = \gamma_i \log (1 - p_i(t))
\end{equation}
The $\gamma_i \in [0, 1]$ denotes the constant systemic fraction, which directly affects the correlation between individual names' spread movements.  $1 - c_i(t)$ is the default probability from the idiosyncratic dynamics, which has to make up the rest of the cumulative hazard according to (\ref{pcon}):
\begin{equation*}
\log (c_i(t)) = (1-\gamma_i) \log(1 - p_i(t)) 
\end{equation*}
This $p_i(x,t)$ specification has an intuitive explanation if we re-write it in the following form:
\[
-\log(1-p_i(x,t)) = \beta_i(t) x - \log (c_i(t))
\]
which is a simple linear regression of a names' cumulative hazard(the LHS) using the common market factor as explanatory variable, and the $-\log(c_i(t))$ is the residual idiosyncratic factor that is specific to the $i$-th name.

This $p_i(x,t)$ specification is convenient because (\ref{p01}) to (\ref{pzz}) are automatically satisfied. (\ref{pzt}) is satisfied as long as the $\beta_i(t)$ is increasing in $t$. A constant $\gamma_i$ in (\ref{hzdsplit}) implies that the $\beta_i(t)$ is not guaranteed to be increasing for all possible $F(x, t)$. Therefore, the choice of either $F(x, t)$ or $\gamma_i$ has to be constrained in order to maintain the monotonicity of the $\beta_i(t)$.

Consider two time periods $t_1 < t_2$ and suppose $f(x, t_1)$ and $\beta_i(t_1)$ are already calibrated to market prices at time $t_1$. With a constant $\gamma_i$, a $\beta_i(t_2) \ge \beta_i(t_1)$ can always be found when the $f(x, t_2)$ is very close to the $f(x, t_1)$ since in the limiting case of $f(x, t_2) = f(x, t_1)$, the $\beta_i(t_2)$ cannot be less than the $\beta_i(t_1)$ given the default probability in (\ref{hzdsplit}) is increasing: $p_i(t_2) \ge p_i(t_1)$. Therefore, the monotonicity of $\beta_i(t)$ can always be enforced by making the $f(x, t_2)$ close to the $f(x, t_1)$.

In a diverse portfolio, the distressed names usually impose more constraints on the choice of $f(x, t_2)$ since their default risk are concentrated in the front end before time $t_1$, and their $p_i(t_2)$ can be very close to $p_i(t_1)$. A constant $\gamma_i$ may force $f(x, t_2)$ to be very close to $f(x, t_1)$ in order to satisfy the monotonicity constraint of $\beta_i(t)$ for the most distressed names in the portfolio, which could undermine the model's ability to calibrate to the index tranches. Therefore for distressed names, it is better to have a time dependent $\gamma_i(t)$ which starts with a low value and increases over time, thus leaving more freedom in the choice of $f(x, t_2)$.  It also makes economic sense for very distressed names to have lower systemic dependencies in the short time horizon.

The $\gamma_i$ factors have to be high ($>80\%$) for the majority of the names in order to obtain good calibration to the index tranches, which suggests that the main risk factor in current market is the systemic risk. For simplicity, $\gamma_i$ is chosen to be 90\% for all names except for very distressed names in this implementation. 

\begin{figure}
\caption{Stochastic Recovery \label{urec}}
\vspace{.25cm}

\center
\begin{minipage}{3in}
\center
\underline{Mean}
\scalebox{.55}{\includegraphics{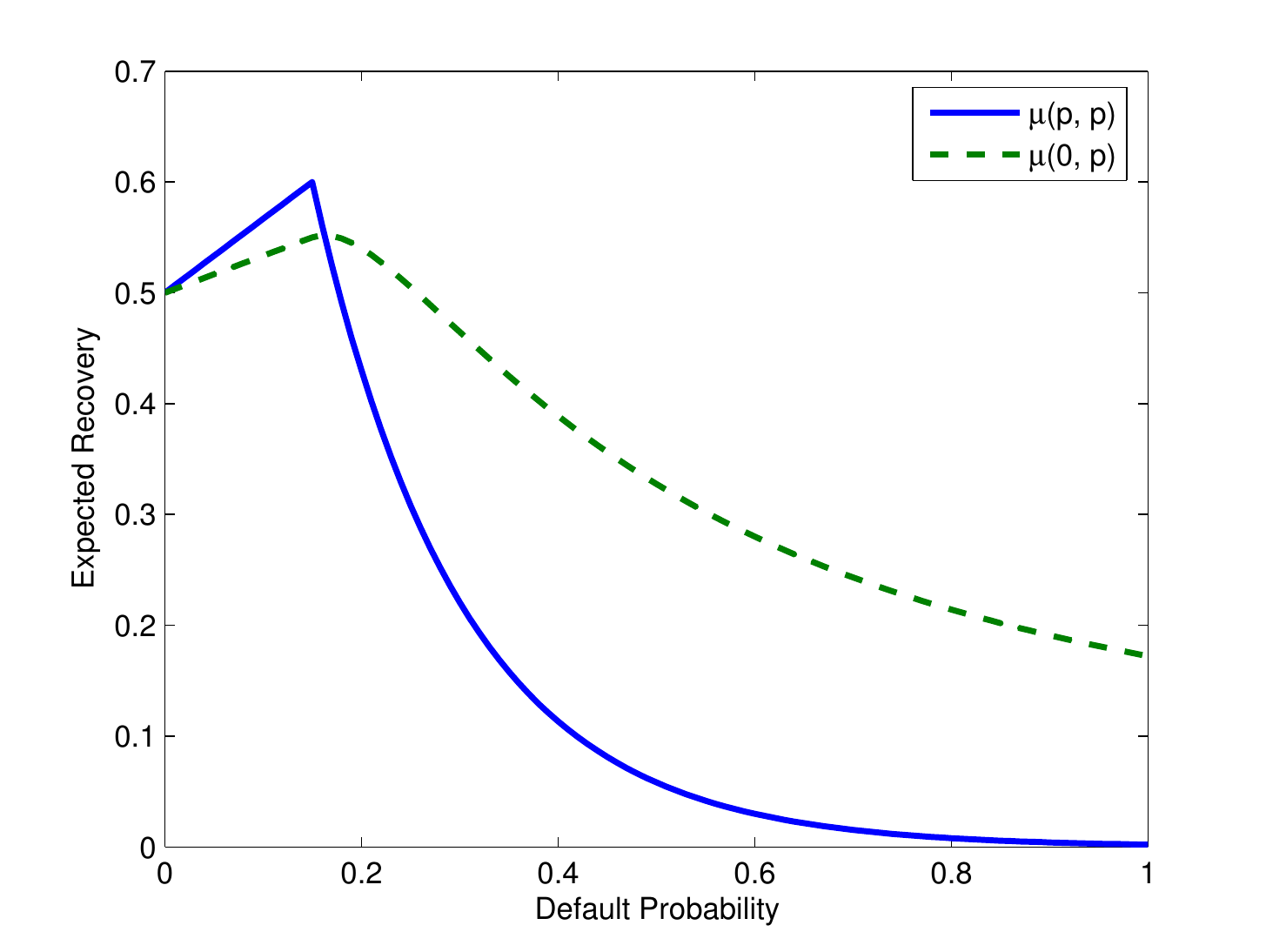}}
\end{minipage}
\begin{minipage}{3in}
\center
\underline{Standard Deviation}
\scalebox{.55}{\includegraphics{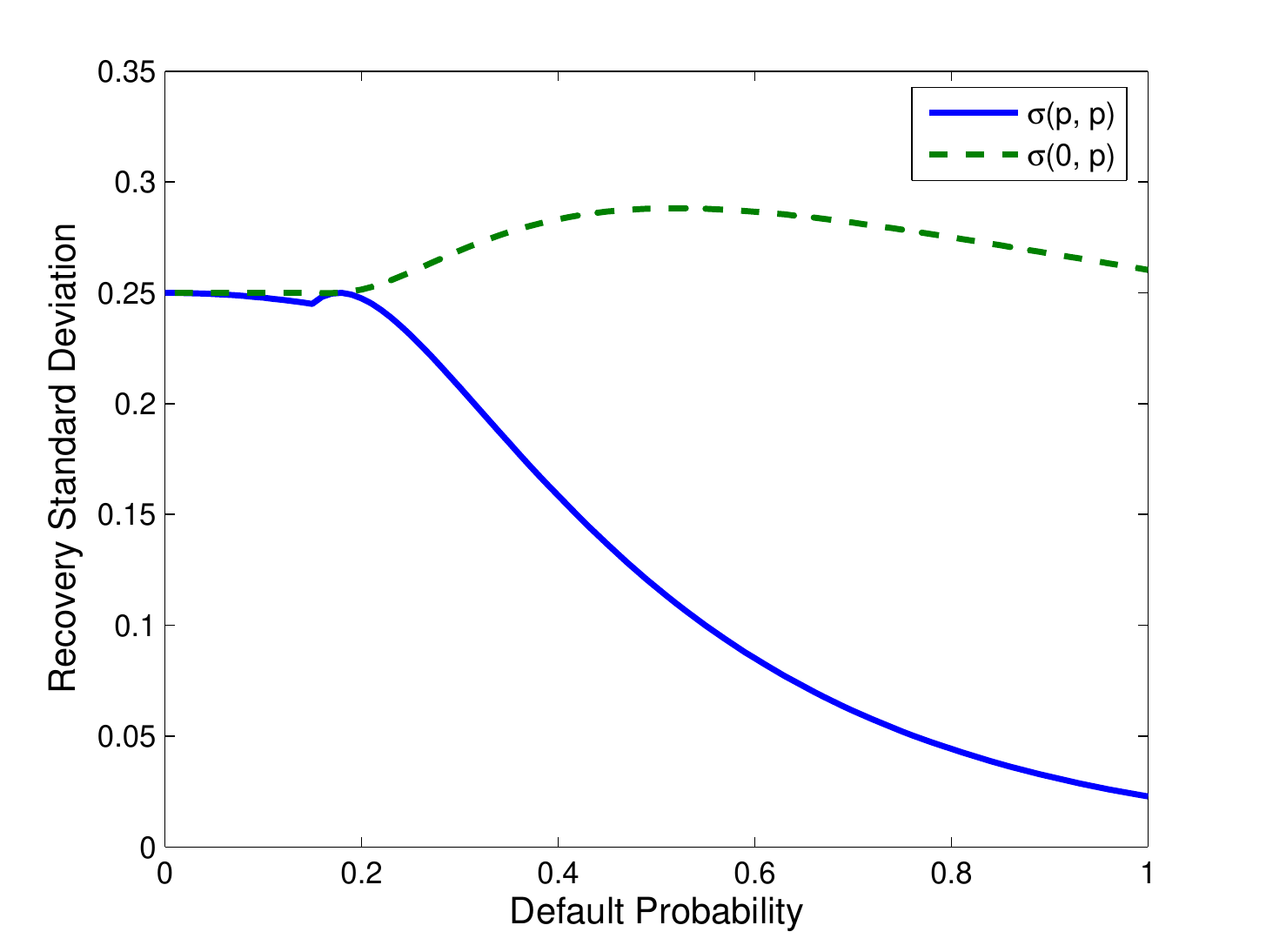}}
\end{minipage}
\end{figure}

\subsection{Stochastic Recovery}
As discussed in section \ref{sr}, only the $\mu_i(p,p)$ and $\sigma_i^2(p,p)$ of the recovery rate need to be specified in order to price CDO tranches consistently.  For simplicity, all credits are assumed to have the same functional form of $\mu(p,p)$ and $\sigma^2(p,p)$.  Figure \ref{urec} shows the mean and standard deviation of the recovery function used in the non-parametric model implementation. The choice of $\mu(p, p)$ function is somewhat arbitrary, its overall trend is chosen to be decreasing in $p$ because it is desirable for the recovery to be lower in the bad states of the market factor. A peak is created in $\mu(p,p)$ at 15\% default probability just to show the ability to create an arbitrary shape of the recovery term structure. The $\sigma^2(p,p)$ is assumed to be a fixed percentage of the maximum variance for the given $\mu(p, p)$: $\sigma^2(p, p) = \alpha \mu(p, p)(1-\mu(p, p))$, where the $\alpha$ is chosen to be 25\% somewhat arbitrarily. If there are observations or views about the variance of a name's recovery rate, the $\alpha$ parameter can be changed to match those. 

The $\mu(p, p)$ function in Figure \ref{urec} is multiplied by a name specific scaling factor to match the individual credits' CDS curve recovery at the 5Y tenor. Since the $\mu, \sigma^2$ are functions of the conditional default probability, the unconditional term recovery rate at time $t$ for name $i$ can be computed by integrating over all the possible market factor values:
\begin{equation}
\label{uncr}
R_i(0, t) = \frac{1}{p_i(t)} \int \mu(0, p_i(x,t)) p_i(x,t) f(x, t) dx
\end{equation}
Even though $\mu(0,p)$ has a strong trend over $p$ as shown in the Figure \ref{urec}, the unconditional recovery rate $R_i(0, t)$ would exhibit a much milder trend over time due to the averaging effects through the integral in (\ref{uncr}). More results about the unconditional recovery rates are shown in the following sections.

The $p_i(x, t)$, $\mu(p, p)$ and $\sigma^2(p, p)$ given in this section are just one example of possible model specifications. There could be many different specifications which are equally valid and effective under the general principles of the Definition \ref{copi}.  

\subsection{Calibration to Index Tranches\label{idxcal}}
In this implementation, the $F(x, t)$ is represented by a non-parametric distribution $\{q_j\}$ at the sampling grid $\{x_j\}$. We first discuss how to calibrate the $\{q_j\}$  to the expected tranche losses (ETL) at a given time horizon, then we discuss the calibration to tranche prices across multiple maturities. 

At a given time horizon $t$, the number of samples in $\{q_j\}$ is generally much greater than the number of tranches, therefore the problem is under-determined: there can be infinitely many distributions that will produce the same input ETL. Some exogenous assumptions on the marginal distribution $\{q_j\}$ are required in order to find a unique solution. We chose to use the maximum entropy method which is well suited to solve this type of under-determined problem in derivative pricing because the resulting distribution from the maximum entropy method contains the least amount of information, thus is the least biased distribution for the given market input. The readers are referred to \cite{luigi} for a detailed discussion on the Maximum Entropy method in CDO tranche calibration. \cite{luigi} applied the maximum entropy method to the loss distribution, the same method can be adopted for the $X_t$ distribution.

The tranche ETLs as computed by \eqref{etl} are linear constraints in the maximum entropy optimization. All the conditional ETLs in \eqref{condetl} have to be computed first in order to apply the linear constraint of \eqref{etl}. However, the conditional ETLs in \eqref{condetl} depend on the $\{q_j\}$ through the conditional default probability $p_i(x, t)$ in (\ref{pcon}); the $\mu_i(0, p_i(x, t))$ and $\sigma_i(0, p_i(x, t))$ of the recovery rate also depends on $\{q_j\}$ through $p_i(x, t)$. To get around this circular dependency between the conditional ETL and $\{q_j\}$, we employed an iterative calibration procedure where at each iteration, the conditional ETLs are first computed from the $\{q_j\}$ of the previous iteration, then the maximum entropy method is used to obtain the $\{q_j\}$ for the next iteration. This iterative procedure works quite well in practice, and usually only a few iterations are needed to converge to a unique solution of $\{q_j\}$ that reproduces the input ETL.

Once we can calibrate the discretely sampled $\{q_j\}$ to the ETL at a given time, we can easily extend the calibration to index tranche prices across multiple maturities by taking advantage of the ``time locality'' property. The calibration at different maturities can be carried out as almost independently except that additional linear constraints are needed in the maximum entropy optimization to ensure the (\ref{phiz}) are met. One technical issue is that the calibration at a single time grid requires the ETL as input, but we only observe the index tranche prics but not the ETLs directly in the market\footnote{The PO contracts on index tranches are not yet liquid enough}. There are two possible ways to address this:
\begin{compactenum}
\item use another model, e.g., base correlation, to extract the ETL surface at each quarterly date and calibrate the model to the full ETL surface. This ensures the maximum consistency to the existing base correlation framework.
\item use an interpolation method on the distributions of the $F(x, t)$ so that the distributions at all quarterly dates can be interpolated from the distributions at the standard maturities of 5Y, 7Y and 10Y. Then we can solve the $\{q_j\}$ and the ETLs at standard index maturities simultaneously during the iterative calibration procedure. At the end of each iteration, we can compute the index tranche prices using the interpolation method on $F(x, t)$, then we can adjust the target ETLs according to the difference between the tranche prices of the current iteration and the input market tranche prices. This adjustment in ETL needs the ratio of change in tranche PV to the change in ETL (ie, $\frac{\partial \textup{PV}}{\partial \textup{ETL}}$), which can be computed from the previous iteration.
\end{compactenum}

\subsection{Spread Dynamics}
Once we calibrated the discrete marginal distribution of $\{q_j\}$ for the common factor process cross the time grid $\{t\}$, we can specify its $JDDT$ by building a discrete Markov chain on $X_t$, and we can further define the $JDDTSF$ by fully specifying the underlying process of $X_t$. 

Two different methods of building the Markov chain on $X_t$ are implemented: co-monotonic and maximum entropy.  A detailed description of these two methods can be found in \cite{losslinker} where both of these methods were applied to the portfolio loss process following the typical top-down approach. The numerical methods in \cite{losslinker} can be applied to the discrete marginal distribution $\{q_j\}$ without modification since the $X_t$ and the loss process has the exact same properties.

Once we constructed the Markov chain, we can further introduce additional systematic factors to complete the dynamics of $X_t$. Even though existing top-down models could be used, we instead propose a very simple extension to the Markov chain of $X_t$. The advantage of this extension is its tractability and efficiency: dynamic instruments whose payoff depends on future spreads can be efficiently priced using a lattice method under this extension to the Markov chains. This extension is also very general, and it does not depends on the details of the non-parametric implementation. We'll discuss this dynamic extension in section \ref{dmc} after presenting some numerical results of the non-parametric model.

\section{Numerical Results\label{numres}}
In this section, some numerical results are presented from the non-parametric implementation of the general framework described in section \ref{imp}. The numerical results presented here uses the market data of CDX-IG9 on the close of Jan. 15, 2009, when the CDX-IG9 index spread is near its historical high. For simplicity of the presentation, we only show the numerical results in ETLs at the standard 5Y, 7Y and 10Y maturuties; it is not difficult to cover the full ETL surface of all quarterly date using either of the two methods described in section \ref{idxcal}. Since the ETL at the maturity is the main driving factor of the tranche prices, the numerical results from the ETLs at the standard maturities can give us great insight to the model properties.

\subsection{Calibration to Index Tranches \label{calidx}}
\begin{figure}
\caption{Model Calibration to CDX-IG9 on Jan. 15, 2009 \label{cdx9cal}}
\center
\vspace{.25cm}

\begin{minipage}{3.2in}
\center
\underline{Market Input ETL}

\vspace{.25cm}
\begin{tabular}{|rr|rrr|}
\hline
 {\bf Att} &  {\bf Det} &   {\bf 5Y} &   {\bf 7Y} &  {\bf 10Y} \\
\hline
     0.0\% &      2.6\% &    83.51\% &    87.23\% &    91.12\% \\

     2.6\% &      6.7\% &    57.22\% &    64.36\% &    71.28\% \\

     6.7\% &      9.8\% &    30.05\% &    41.47\% &    54.94\% \\

     9.8\% &     14.9\% &    18.02\% &    26.07\% &    36.49\% \\

    14.9\% &     30.3\% &     4.87\% &     7.20\% &    10.57\% \\

    30.3\% &     61.0\% &     4.05\% &     6.24\% &     8.54\% \\
\hline
     0.0\% &    100.0\% &     8.72\% &    10.96\% &    13.47\% \\
\hline
\end{tabular}  
\end{minipage}
\begin{minipage}{3.2in}
\center
\underline{ETL from Model Calibration}

\vspace{.25cm}
\begin{tabular}{|rr|rrr|}
\hline
 {\bf Att} &  {\bf Det} &   {\bf 5Y} &   {\bf 7Y} &  {\bf 10Y} \\
\hline
     0.0\% &      2.6\% &    83.54\% &    87.11\% &    90.62\% \\

     2.6\% &      6.7\% &    57.27\% &    64.10\% &    70.43\% \\

     6.7\% &      9.8\% &    30.16\% &    41.42\% &    54.52\% \\

     9.8\% &     14.9\% &    18.03\% &    25.83\% &    35.57\% \\

    14.9\% &     30.3\% &     5.02\% &     7.73\% &    11.01\% \\

    30.3\% &     61.0\% &     3.98\% &     5.74\% &     7.42\% \\
\hline
     0.0\% &    100.0\% &     8.72\% &    10.96\% &    13.47\% \\
\hline
\end{tabular}  
\end{minipage}
\end{figure}

\begin{figure}
\center
\caption{Calibrated $F(x,t)$ \label{xcal}}
\center
\scalebox{.6}{\includegraphics{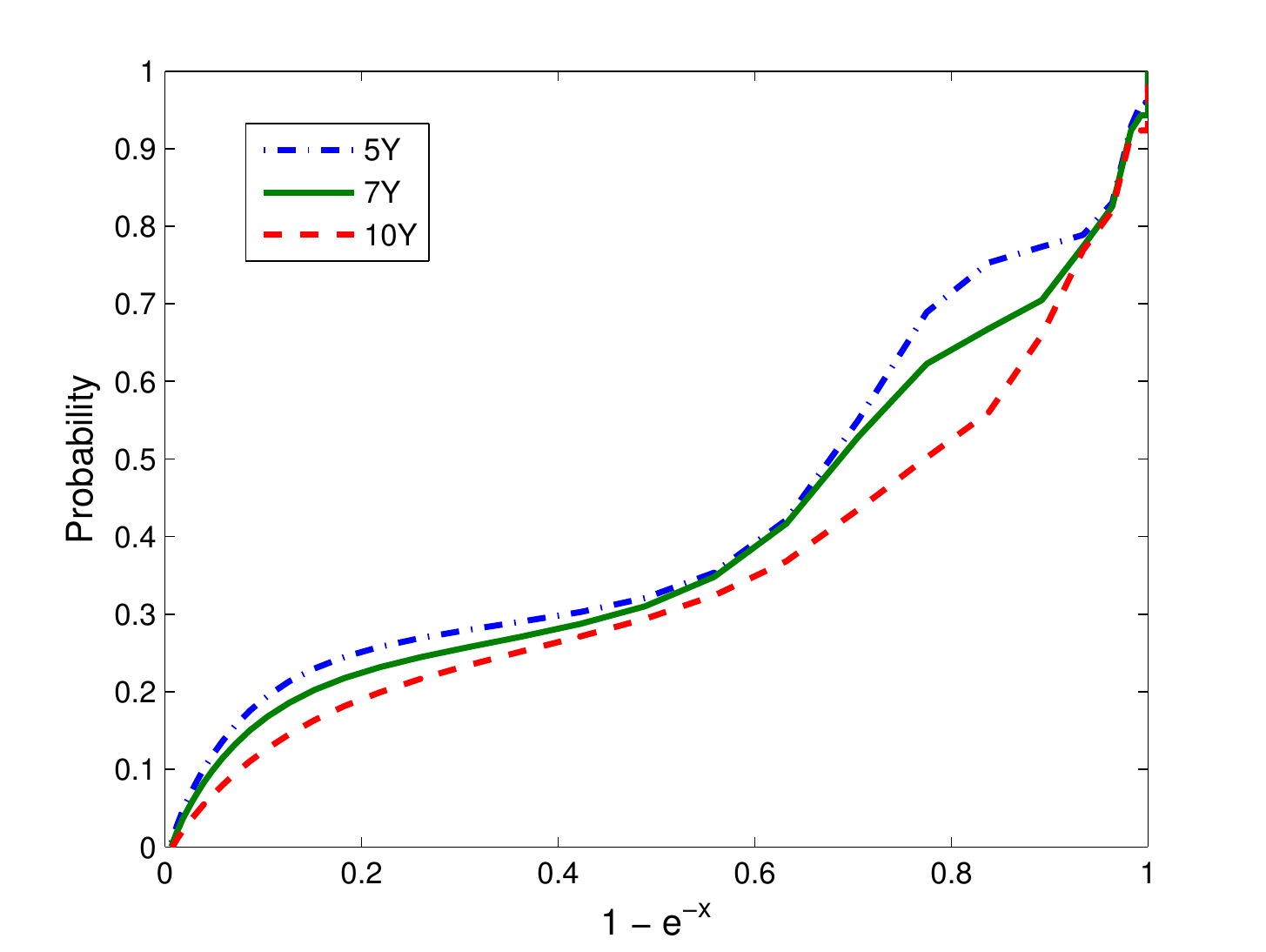}}
\end{figure}

Following the method desribed in section \ref{idxcal}, the non-parametric implementation is calibrated to the ETLs of CDX-IG9 index as of the close of Jan. 15, 2009. Figure \ref{cdx9cal} shows the input ETLs from the tranche market and the model calibration results. The market ETL inputs are extracted from a standard base correlation model. The non-parametric model is able to calibrate quite closely to the input ETL across the three maturities. Figure \ref{xcal} showed the calibrated cumulative distribution function $F(x,t)$ at 5Y, 7Y and 10Y. The constraint (\ref{phiz}) is built into the bootstrap process so that the resulting marginal distributions are compatible with an increasing process. It is visually obvious that the calibrated $F(x,t)$ indeed satisfy (\ref{phiz}) since the three CDF curves never cross each other. The iterative calibration procedure described in \ref{idxcal} is very fast, it only takes a few seconds on a regular PC to calibrate the model to the 5Y, 7Y and 10Y ETLs.

Since the model expected recovery matches the CDS curve recovery at only the 5Y maturity, the single name default probabilities at the 7Y and 10Y tenors are adjusted accordingly to preserve the expected loss\footnote{We can also choose to match the CDS spread or upfront instead of the expected loss. We choose to match the expected loss because the inputs to calibration are expected tranche losses, and we want to preserve the portfolio expected loss. Matching CDS expected loss results in very similar CDS spreads or upfronts as the inputs since the PV01 differences due to recovery changes are normally very limited.} of the input CDS curves. The calibration results showed that the expected portfolio losses of the 0-100\% tranche are exactly preserved at all the maturities.

\subsection{Implied Recovery Rate Term Structure}
\begin{figure}
\caption{Expected Recovery Change \label{7YRec}}
\center
\vspace{.25cm}

\begin{minipage}{3in}
\center
\underline{7Y}
\scalebox{.55}{\includegraphics{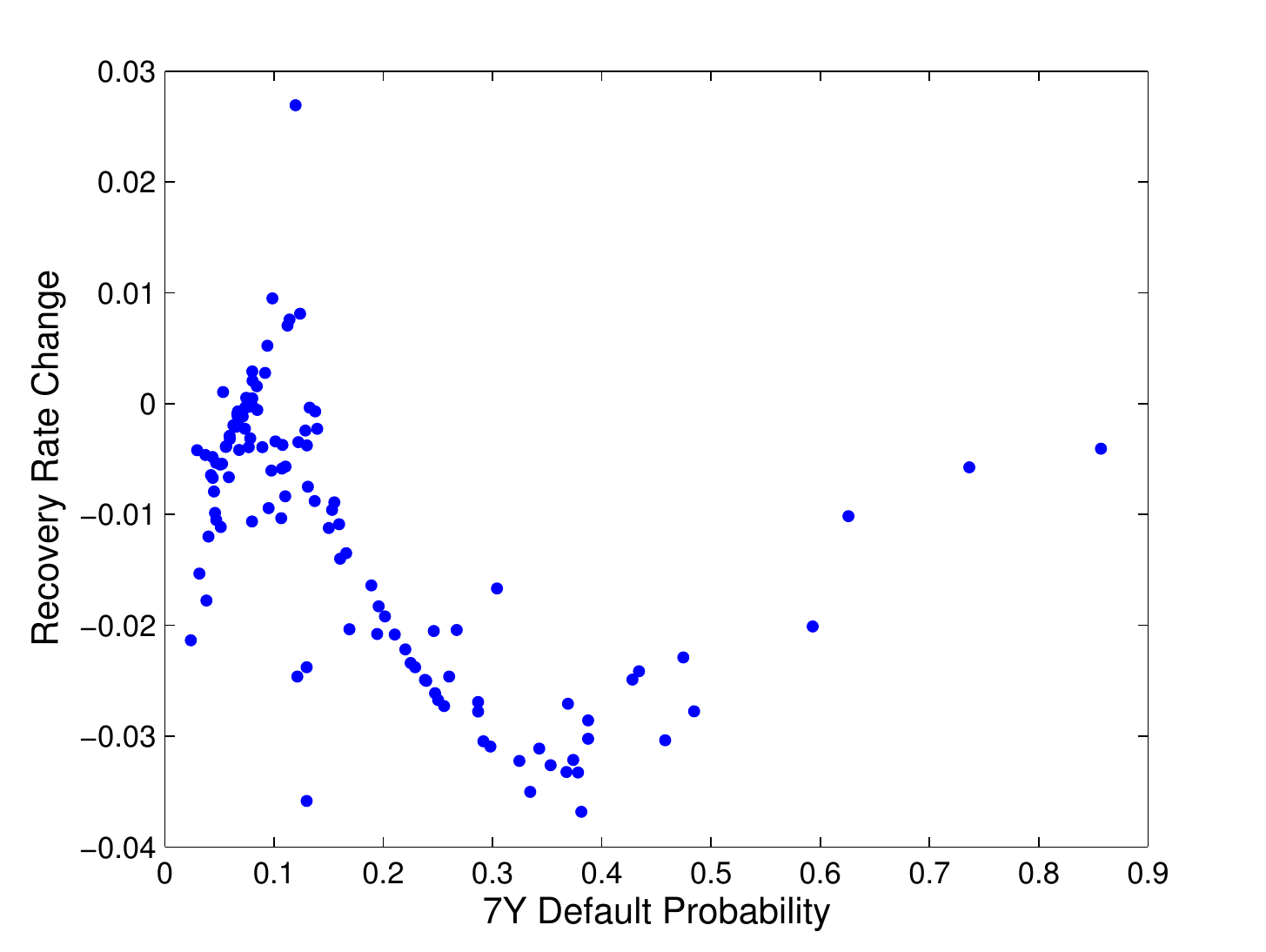}}
\end{minipage}
\begin{minipage}{3in}
\center
\underline{10Y}
\scalebox{.55}{\includegraphics{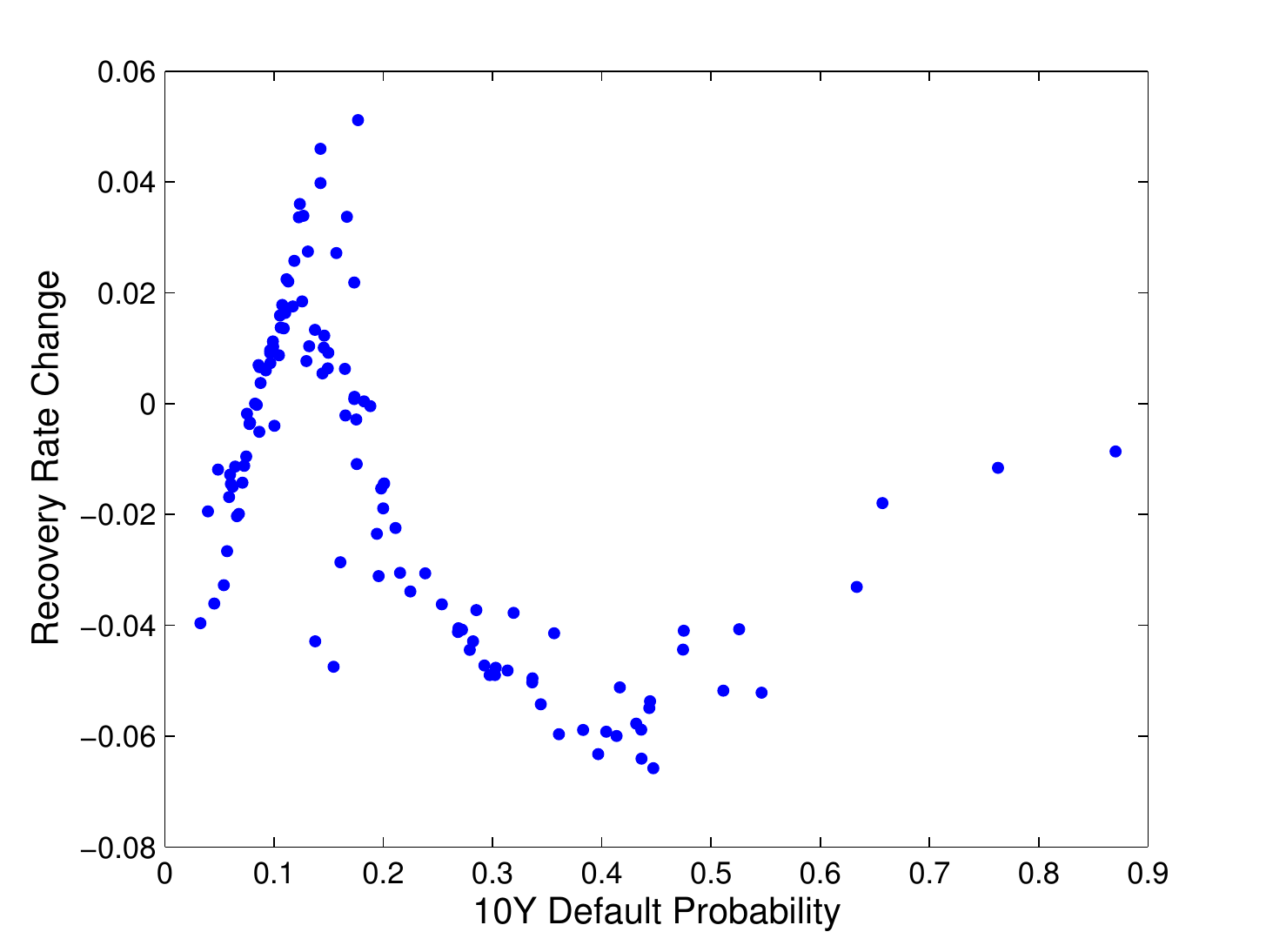}}
\end{minipage}
\end{figure}

As the most common practice, traders only mark a single recovery value for a CDS curve, which we refer to as the ``curve recovery''. The calibrated model matches the CDS curve recovery exactly at the 5Y tenor, but not at the 7Y and 10Y. Figure \ref{7YRec} showed the scatter plots of the difference between the 7Y and 10Y model implied recoveries from the curve recoveries for all the 122 names in the CDX-IG9 portfolio. The horizontal axis is the default probability at the corresponding tenor.  Figure \ref{7YRec} showed that the model expected recoveries at 7Y and 10Y only differ by a few percentage points at most from the curve recoveries. Given that the recovery locks are only traded for distressed names at very short maturity with wide bid/offer, the small deviation from curve recovery is not a problem in practice.

\subsection{Monte Carlo Simulation\label{secmc}}

\begin{figure}
\caption{Monte Carlo Simulation of Tranche Loss \label{mc}}
\center
\vspace{.25cm}

\begin{minipage}{3.2in}
\center
\underline{Co-monotonic Markov Chain}
\vspace{.25cm}

\begin{tabular}{|rr|rrr|}
\hline
 {\bf Att} &  {\bf Det} &   {\bf 5Y} &   {\bf 7Y} &  {\bf 10Y} \\
\hline
     0.0\% &      2.6\% &    83.59\% &    87.15\% &    90.64\% \\

     2.6\% &      6.7\% &    57.18\% &    64.10\% &    70.46\% \\

     6.7\% &      9.8\% &    30.12\% &    41.35\% &    54.48\% \\

     9.8\% &     14.9\% &    18.02\% &    25.81\% &    35.49\% \\

    14.9\% &     30.3\% &     5.02\% &     7.73\% &    11.01\% \\

    30.3\% &     61.0\% &     3.97\% &     5.73\% &     7.40\% \\
\hline
     0.0\% &    100.0\% &     8.71\% &    10.95\% &    13.46\% \\
\hline
\end{tabular}  
\end{minipage}
\begin{minipage}{3.2in}
\center
\underline{Maximum Entropy Markov Chain}
\vspace{.25cm}

\begin{tabular}{|rr|rrr|}
\hline
 {\bf Att} &  {\bf Det} &   {\bf 5Y} &   {\bf 7Y} &  {\bf 10Y} \\
\hline
     0.0\% &      2.6\% &    83.54\% &    87.08\% &    90.62\% \\

     2.6\% &      6.7\% &    57.18\% &    64.08\% &    70.46\% \\

     6.7\% &      9.8\% &    30.16\% &    41.36\% &    54.47\% \\

     9.8\% &     14.9\% &    18.00\% &    25.84\% &    35.52\% \\

    14.9\% &     30.3\% &     4.97\% &     7.70\% &    11.00\% \\

    30.3\% &     61.0\% &     3.93\% &     5.72\% &     7.39\% \\
\hline
     0.0\% &    100.0\% &     8.68\% &    10.94\% &    13.45\% \\
\hline
\end{tabular}  
\end{minipage}
\end{figure}

A simple Monte Carlo simulation of default time is implemented in this section to verify the consistency and correctness of the proposed modelling framework.  A Monte Carlo simulation of default times can also be used to price exotic instruments that only depends on the $JDDT$ but not the $JDDTSF$. According to Proposition \ref{info}, a Markov chain of $X_t$ is needed in order to simulate the default times according to the $JDDT$.
 
Figure \ref{mc} showed the simulated ETLs at the three maturities from drawing 1,000,000 independent default time and recovery paths from both of the co-monotonic and maximum entropy Markov chains. The default time and recovery paths are drawn using the following steps:
\begin{compactenum}
\item Draw a full path of $X_t$ over time from the Markov chain.
\item Use the $p_i(x,t)$ function to compute the conditional default probability term structures of all the underlying names for the given path of $X_t$.
\item For each name, draw an independent uniform random number $d_i$ which represents the conditional default probability. $d_i$ is then used to determine the default period of the corresponding name according to the conditional default probability term structure. 
\item For each name defaulted before the final maturity (10Y), compute its spot recovery mean and variance $\mu_i(d_i, d_i)$, $\sigma^2_i(d_i, d_i)$. 
\item Draw an independent recovery rate for any defaulted name from a two point distribution whose mean and variance are given by the $\mu_i(d_i, d_i)$, $\sigma^2_i(d_i, d_i)$.
\end{compactenum}
After drawing the default time and recovery path, the tranche losses at all tenors are computed from the same default time and recovery path to ensure full consistency across all maturities. Then the tranche losses from these independent default time and recovery paths are averaged to produce the ETL.

The simulated ETLs from the two Markov chains are very close to each other, which is expected since they have identical $\{JDDI(t)\}$ by construction.  Both of the simulated ETLs are very close to the semi-analytical calibration results shown in Figure \ref{cdx9cal}, where the normal approximation is used to build the conditional loss distribution. The maximum difference in the ETL between the Monte Carlo simulation and the semi-analytical pricing with normal approximation is less than 0.1\%. The ETL difference of this magnitude is clearly negligible for practical purposes.  It is also verified that a different spot recovery rate distribution, such as the beta distribution, produces very similar results to those in Figure \ref{mc}, as long as the $\mu(p,p)$ and $\sigma^2(p,p)$ of the recovery rate are matched.

\begin{figure}
\caption{Temporal Correlation of Incremental Portfolio Losses \label{corrloss}}
\center
\vspace{.25cm}

\begin{minipage}{3in}
\center
\underline{Co-monotonic Markov Chain}

\vspace{.25cm}
\begin{tabular}{|r|rrr|}
\hline
 {\bf -}  &  {\bf 0-5Y} &   {\bf 5Y-7Y} &   {\bf 7Y-10Y} \\
\hline
 {\bf 0-5Y}  &      1 &    .5027 &    .4887  \\

 {\bf 5-7Y}  &      .5027 &    1 &    .2109  \\

 {\bf 7-10Y} &      .4887 &    .2109 &    1  \\
\hline
\end{tabular}  
\end{minipage}
\begin{minipage}{3in}
\center
\underline{Maximum Entropy Markov Chain}

\vspace{.25cm}
\begin{tabular}{|r|rrr|}
\hline
 {\bf -}  &  {\bf 0-5Y} &   {\bf 5Y-7Y} &   {\bf 7Y-10Y} \\
\hline
 {\bf 0-5Y}  &      1 &    .4199 &    .3936  \\

 {\bf 5-7Y}  &      .4199 &    1 &    .1227  \\

 {\bf 7-10Y} &      .3936 &    .1227 &    1  \\
\hline
\end{tabular}  
\end{minipage}
\end{figure}

However, the two Markov chains lead to very different $JDDT$s. Figure \ref{corrloss} shows the correlation matrix between the simulated incremental portfolio losses in the three periods (0-5Y, 5Y-7Y and 7Y-10Y), conditioned on the portfolio loss before 5Y is less than 10\%. It is evident that the temporal loss correlation from the co-monotonic Markov chain is much stronger than that of the maximum entropy Markov chain. The temporal loss correlation is a critical factor in pricing exotic correlation instruments such as forward-starting tranche and loss-triggered LSS. This example showed that top-down methods can be applied to change the $JDDT$ while preserving the calibrated $\{JDDI(t)\}$ due to Proposition \ref{info}.

\begin{figure}
\center
\caption{Simulated Single Name Expected Loss\label{sn}}
\center
\vspace{.25cm}

\begin{minipage}{3in}
\center
\underline{Co-monotonic Markov Chain}
\scalebox{.55}{\includegraphics{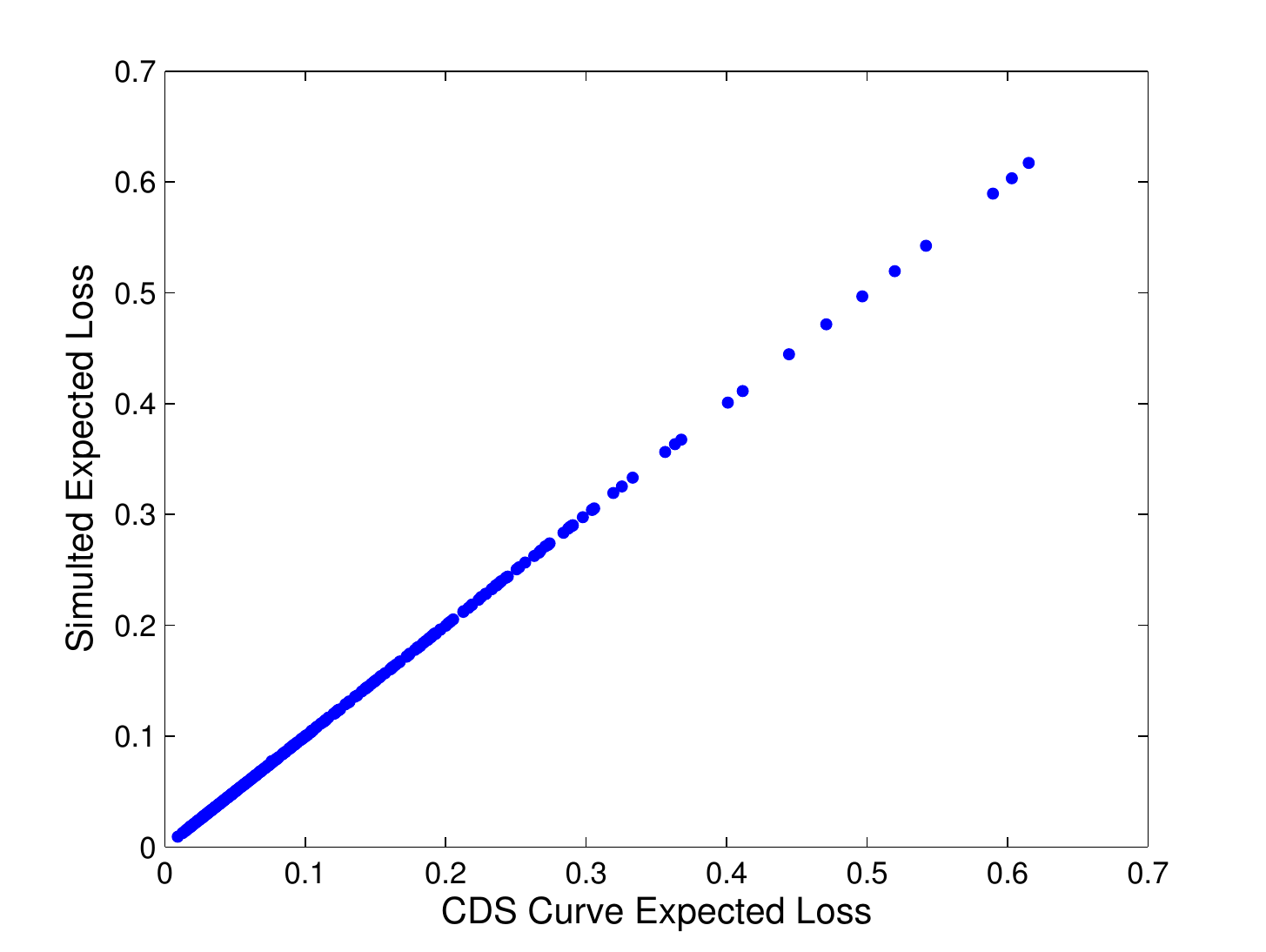}}
\end{minipage}
\begin{minipage}{3in}
\center
\underline{Maximum Entropy Markov Chain}
\scalebox{.55}{\includegraphics{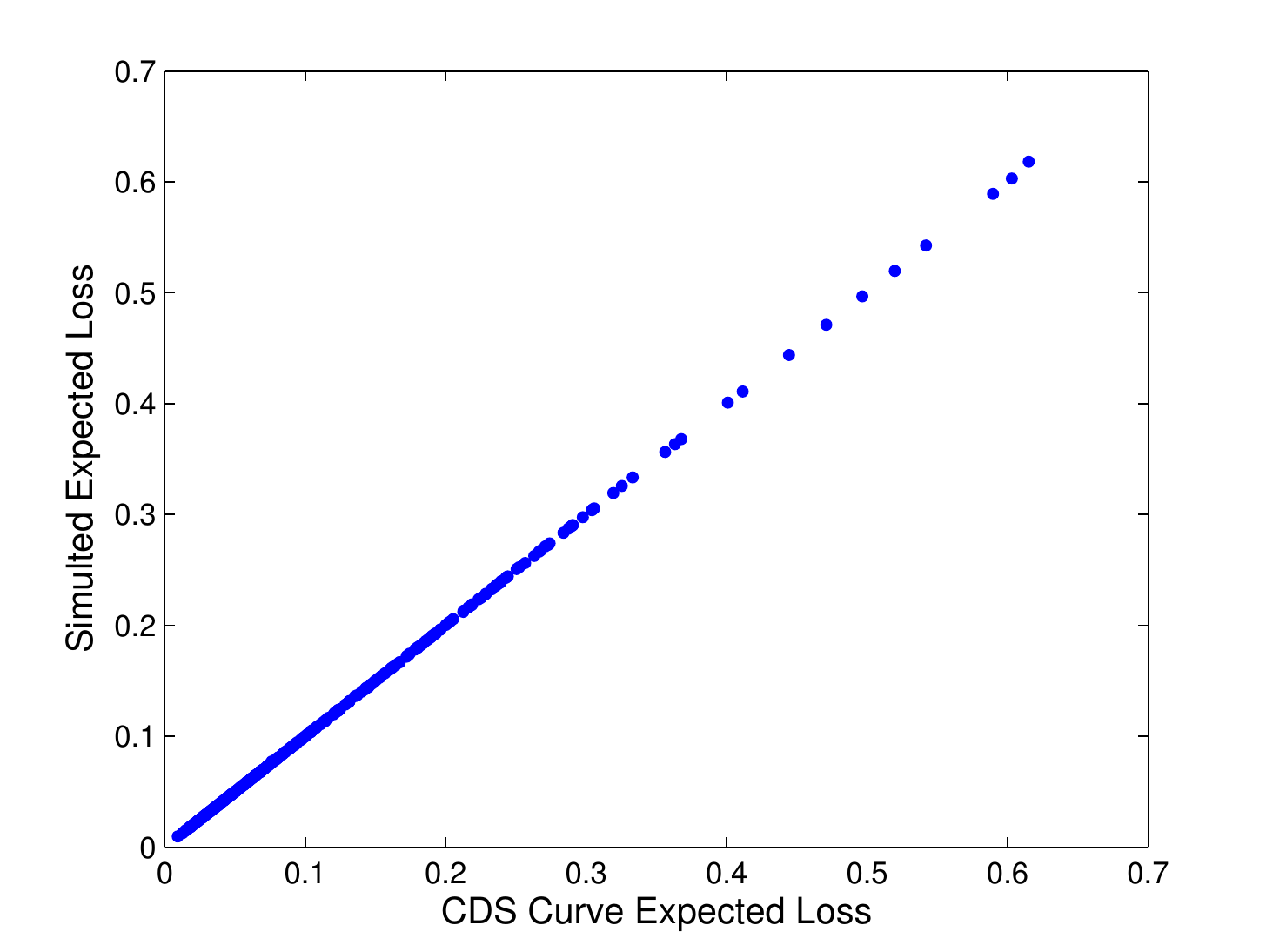}}
\end{minipage}
\end{figure}

Figure \ref{sn} showed two scatter plots of the simulated vs. the CDS curve expected losses for all 122 underlying names over the three maturities. All the dots in Figure \ref{sn} are perfectly aligned along the diagonal line in this scatter plot, which showed that the Monte Carlo simulation correctly preserves all the single names' expected losses across all three maturities.

\section{Conditional Markov Chain\label{dmc}}
Exotic instruments whose payoff depends on future spreads, such as tranche options, can be priced from the $JDDTSF$ and the idiosyncratic dynamics. The idiosyncratic dynamics is easy to handle because they are independent of other factors by definition. Therefore, we can price these spread-dependent exotic instruments if we can fully specify the dynamics of $X_t$ and the $JDDTSF$. Since the Markov chain of $X_t$ is specified in the second step of the progressive calibration procedure in Table \ref{calprog}, it would be the most convenient numerically if we can specify the $X_t$ dynamics in the third step to be consistent with the Markov chain from the second step. In this section, we propose a conditional Markov chain method that fully specifies the dynamics of $X_t$ while maintaining consistency with its Markov chain, thus allowing the $JDDTFS$ to be changed without changing the $JDDT$ and $\{JDDI(t)\}$.

Suppose the time is discretely sampled by $\{t\}$ and the market factor process $X_t$ is discretely sampled by a fixed grid of $\{x_j\}$; we denote the discrete Markov Chain of $X_t$ as $\mathbb{P}\{X_{t+1} \le x | X_t \}$, which is the probability of $X_{t+1} \le x$ for any $x$ conditioned on the value of $X_t$. We assume a simple Ornstein-Uhlenbeck driver process $y_s$ exists for the Markov chain:
\[
dy_s = \kappa(y_s - \bar{y}) ds + v dW_s
\]
The OU process is parameterized by its long run mean $\bar{y}$, mean reversion coefficient $\kappa$ and volatility $v$. We've chosen the simple OU process because its $y_t$ distribution is Gaussian, and its mean and variance are easy to compute:
\begin{align*}
\mu_t &= \mathbb{E}[y_t] = y_0e^{-\kappa t} + \bar{y}(1-e^{-\kappa t}) \\
\sigma^2_t &= \textup{Var}[y_t] = \frac{v^2}{2 \kappa}(1-e^{-2\kappa t})
\end{align*}
where $y_0$ is the initial value of $y_s$. We then define a $z_t$ process from the driver process:
\begin{equation*}
z_{t+1} = \beta \frac{y_t - \mu_t}{\sigma_t}   + \sqrt{1-\beta^2} e_{t+1}
\end{equation*}
where $e_{t+1}$ is an independent standard normal random variable; therefore $z_{t+1}$ is also standard normal. The $z_{t+1}$ determines the outcome of the transition from $X_t$ to $X_{t+1}$. Similar to a Gaussian Copula, we can define a threshold $c_j(X_t)$ for each possible outcome of $X_{t+1} = x_j$:
\begin{equation*}
\mathbb{P}\{z_{t+1} < c_j(X_t) | X_t \} = \mathbb{P}\{X_{t+1} \le x_j | X_t \}
\end{equation*}
Similar to the conditional default probability in the standard Gaussian Copula, we can then compute the transition probability conditioned on the value of $y_t$ and $X_t$:
\begin{equation}
\label{condtd}
\mathbb{P}\{X_{t+1} \le x | X_t, y_t \} = \mathbb{P}\{z_{t+1} < c_j(X_t) | X_t, y_t\} = \Phi(\frac{c_j(X_t) - \beta \frac{y_t-\mu_t}{\sigma_t}}{\sqrt{1-\beta^2}})
\end{equation}
Note that even though $y_{t}$ is standard normal unconditionally, its distribution is generally not standard normal conditioned on $X_t$, therefore, the threshold $c_j(X_t)$ has to be determined by the following 
relationship:
\begin{equation}
\label{threshold}
\mathbb{P}\{X_{t+1} \le x | X_t \} = \mathbb{E}[\mathbb{P}\{X_{t+1} \le x | X_t, y_t \} |X_t] = \int \Phi(\frac{c_j(X_t) - \beta \frac{y_t-\mu_t}{\sigma_t}}{\sqrt{1-\beta^2}}) f(y_t | X_t) dy
\end{equation}
where $f(y_t|X_t)$ is the distribution of $y_t$ conditioned on the value of $X_t$. 

The $X_t$ is the common economic factor that encapsulates the overall health of the economy. In this simple specification, the $z_{t+1}$ determines the outcome of the $X_{t+1}$ from $X_t$; the $z_t$ process 
can be viewed as the underlying economic factors that drives the trend of overall economic movements. The $z_{t+1}$ consists of two parts, a time-persisting $\frac{y_t - \mu_t}{\sigma_t}$ that affects multiple periods, 
and a random shock $e_{t+1}$ that only affect a single period from $t$ to $t+1$. The $\beta$ parameter controls the mixture of these two contributing factors. The $y_t$ can be viewed as the slow-moving market wide economic 
forces such as overall production and consumption etc., and $e_t$ are random shocks such as natural disaster or unpredictable geopolitical events. The $y_t$ process is mean-reverting to capture the overall economic 
cycles. Under the conditional Markov chain, the market filtration $\mathcal{F}_t$ includes $X_t, y_t$ and the realized defaults.

The $\beta$ parameter is very important in this specification. The higher the $\beta$, the more information we can infer about the future distribution of $X_t$ by observing $y_t$. In the limiting case of $\beta = 0$, $y_t$ process gives no additional information. Everything else equal, a higher $\beta$ will cause the tranche prices at $t$ to be more volatile because they are more sensitive to the value of $y_t$, which leads to higher value of tranche options. Thus we can use the $\beta$ parameter to calibrate the model to tranche options if their prices become observable. The $\beta$ parameter can also be made time- and $X_t$-dependent to match option prices across time and capital structure. Therefore, this simple conditional Markov chain extension allows straight-forward calibration to tranche option prices cross time and capital structure. 

In this simple specification, the transition outcomes from all the values of $X_t$ are controlled by the same $y_t$ process. We could use different $y_t$ processes for different values of $X_t$, but there is no obvious 
economic justification or practical benefits of that; therefore we choose to use same $y_t$ for all $X_t$ for simplicity. 

The advantage of this simple specification is its tractability and flexibility. The conditional transition probability for a single time period is explicitly given by \eqref{condtd}; and the conditional transition probability is fully consistent with the unconditional transition probability of the Markov Chain by construction. A two-dimensional lattice of $(X_t, y_t)$ can be built numerically for multiple time periods, and various dynamic payoffs can be priced via backwards induction on the $(X_t, y_t)$ lattice. Since the 2-D lattice accurately tracks the conditional distribution $f(y_t|X_t)$, the threshold $c_j(X_t)$ can be easily computed from \eqref{threshold} at each time step. 

The lattice pricing technique suggested here has some similarity to the lattice method in \cite{ml}. The numerical methods to build the two-dimensional lattice here is much simpler than those in \cite{ml} since the full 2-D lattice can be built using the exact formulas in \eqref{condtd} and \eqref{threshold}; while the 2-D lattice in \cite{ml} is constructed via an approximation to a partial integral differential equation (PIDE), which is non-trivial numerically. The root searching in \eqref{threshold} is the only time-consuming part of building the 2-D lattice, which has a similar 
order of complexity as a single CDO pricing in the Random Factor Loading model described in \cite{rfl}. Therefore, the numerical construction of the 2-dimensional lattice and the subsequent pricing of dynamic payoff 
should take a similar amount of time as the pricing of a single CDO tranche under the RFL model, which is fast enough for practical pricing and risk management purposes.

To illustrate the lattice pricing method, we constructed the 2-D $(X_t, y_t)$ lattice on the maximum entropy Markov chain built from section \ref{secmc} with the follow parameters $\kappa = .05, \bar{y} = y_0 = v = 1$. The 2-D lattice is then used to price the 5Y to 10Y European tranche options\footnote{The 5Y and 10Y standard maturity for IG9 are Dec 20, 2012 and Dec 20, 2017}, where the holder has the right (not the obligation) to buy protection on a 10Y zero-coupon CDX-IG9 tranche at 5Y with fixed strike prices equal to the 10Y ETL. For simplicity, we ignored all the discounting factors\footnote{The effects of discounting is roughly a constant multiplying factor on all the option prices}. It only takes a few seconds to build the 2-D lattice and price all the tranche options in Figure \ref{atmo} on a regular PC. As expected, the higher $\beta$ values result in higher tranche option prices in this example, but it is noticeable that option values on senior tranches are generally not very sensitive to the $\beta$. The choice of Markov chain also have a strong impact on the tranche option values, for example, the co-monotonic Markov chain would results in higher tranche option valuation because the future market factor distribution is more predictable from its current value under a co-monotonic Markov chain.

\begin{figure}
\caption{CDX-IG9 5Y to 10Y Tranche Option Prices\label{atmo}}
\center
\vspace{.25cm}

\begin{tabular}{|rr|rrrrr|}
\hline
\multicolumn{ 2}{|c}{{\bf IG9 Tranches}} & \multicolumn{ 5}{|c|}{{\bf $\beta$ Values}}     \\
\hline
{\bf Att} & {\bf Det} & {\bf 0\%} & {\bf 25\%} & {\bf 50\%} & {\bf 75\%} & {\bf 100\%} \\
\hline
0.0\% & 2.6\% & 6.80\% & 6.80\% & 6.82\% & 6.86\% & 6.92\% \\
2.6\% & 6.7\% & 17.76\% & 17.75\% & 17.75\% & 17.79\% & 17.91\% \\
6.7\% & 9.8\% & 18.92\% & 19.05\% & 19.32\% & 19.63\% & 20.05\% \\
9.8\% & 14.9\% & 15.97\% & 16.30\% & 16.80\% & 17.34\% & 17.85\% \\
14.9\% & 30.3\% & 8.13\% & 8.13\% & 8.13\% & 8.13\% & 8.13\% \\
30.3\% & 61.0\% & 6.73\% & 6.73\% & 6.73\% & 6.75\% & 6.81\% \\
60.0\% & 100.0\% & 1.73\% & 1.73\% & 1.73\% & 1.74\% & 1.77\% \\
\hline
\end{tabular}

\end{figure}

Similar to \cite{ml}, the 2-D lattice does not keep track of realized defaults or idiosyncratic default factors for numerical tractability. Ignoring this information generally leads to sub-optimal exercise of the option, therefore the tranche option prices from the lattice method is actually a lower bound rather than the exact price in the strict sense. However, since the realized defaults and the $X_t$ process are highly correlated in this model setup, and the idiosyncratic dynamics generally contribute very little as discussed in section \ref{condfun}; the resulting prices from the lattice method should be a very close lower bound as very little new information can be added by the realized loss and idiosyncratic factors. The exact pricing rather than a close lower bound can be obtained by a least-square Monte Carlo simulation as described in \cite{ls}. Both the realized loss and idiosyncratic dynamics can be tracked accurately within the Monte Carlo simulation. The Monte Carlo simulation is very useful for checking the accuracy of the lattice implementation, and for quick turn around of exotic structures. However, in most practical situations, the lattice method is preferred since it is much faster and allows easy computation of deltas and risks.
 
In \cite{ml}, the $y_t$ process determines both the unconditional transition probability and conditional transition probability of the $X_t$. In this setup, the $y_t$ process only controls the conditional transition probability of the $X_t$, and the unconditional transition probability of $X_t$ can be either calibrated to relevant market information, or can be specified exogenously as in the example of the Maximum Entropy Markov Chain. Our approach is more flexible because it allows the users to choose the unconditional transition rate of 
$X_t$ and $JDDT$ directly. The model implied tranche price and loss distribution are not affected by the choice of the Markov chain because the $JDDI(t)$ remains invariant. Whereas in \cite{ml}, there is no easy way to adjust the unconditional Markov chain or the $JDDT$ directly since the the $y_t$ process itself is calibrated to index tranche market, and changing its parameter would change the loss distribution and tranche prices. Also, 
the $X_t$ process in \cite{ml} is always continuous because it is an integration of the $y_t$ process, while the $X_t$ in our specification admits large jumps. Therefore, our specification is more general than \cite{ml}, and it can capture a wider variety of possible market spread dynamics.

The affine jump diffusion (AJD) process is a very popular choice recently in building the bottom-up dynamic correlation models. In AJD models, the jump is usually modelled as independent Poisson jumps with a deterministic hazard rate for tractability, as in \cite{ml}.  In such an AJD model, the senior tranches only suffer losses once a large jump arrives. Since a Poisson process is memoryless, the probability of large jumps does not depends on any systemic state variables in the market filtration. Therefore the senior tranche's expected loss and spreads tend to exhibit very low volatility in such an AJD dynamic model. In the proposed conditional Markov chain, the $X_t$ process can have large jumps, and the probability of large jump arrivals depends on the current value of $y_t$, therefore this modelling framework can produce high senior tranche volatility as observed in the recent market.

Even though it is quite easy to construct other alternative specifications of the $X_t$ process following the general framework described in this paper, the conditional Markov Chain method described here has the advantage of being very simple, tractable and fast. Therefore it could be a practical solution to price and manage exotic correlation products.

\section{Conclusion}
This paper proposed a tractable and consistent stochastic recovery specification, and a very generic dynamic correlation modelling framework that combines the best features of the top-down and bottom-up approaches: it is fully consistent with all single name information and admits very rich and flexible spread dynamics. The modelling framework is equipped with the important ``time locality'' property, which allows easy and accurate calibration to the index tranche prices across multiple maturities. Calibration to the index tranches across multiple maturities in a consistent model has been a very difficult modelling problem, and the ``time locality'' property is the key to address it. 

The Property \ref{info} of the proposed modelling framework allows us to calibrate the model progressively (as in table \ref{calprog}) to different  types of market instruments. Vanilla instruments, such as CDO tranches, can be efficiently priced using the semi-analytical method with normal approximation. The conditional Markov chain in section \ref{dmc} is a very simple and fast method to price dynamic instruments, such as tranche options.  Therefore, this modelling framework can cover a wide variety of credit instruments and can be very useful in practice.

\bibliographystyle{apsr}
\bibliography{bd}
\end{document}